\documentclass[runningheads]{comsis2}

\def\journalissue{Computer Science and Information Systems 00(0):0000--0000}
\def\paperidnum{https://doi.org/10.2298/CSIS123456789X}
\usepackage{hyperref}
\usepackage{verbatim}
\usepackage{graphicx}

\usepackage{epsfig}
\usepackage{amssymb}
\usepackage{bm}
\usepackage{enumerate}

\usepackage{amsmath}
\usepackage{epstopdf}
\usepackage{subfigure}
\usepackage{multirow}
\usepackage{algorithm}
\usepackage{algorithmic}

\title{Efficient algorithms for collecting the statistics of large-scale IP address data
\footnote{This work was supported in part by the Guangdong Province Science and Technology Innovation Strategy Special Fund under Grant PDJH2024B648, in part by the Guangdong Province Characteristic Innovation Project for Normal Universities under Grant 2023KTSCX338,  and in part by the Province Ordinary Higher Education Engineering Technology Research (Development) Center under Grant 2024GCZX028.}}

\titlerunning{Efficient algorithms for ...}
\author{Hui Liu\inst{1} and Yi Cao\inst{1}\thanks{Corresponding author.} and Zehan Cai \inst{1} \and Hua Mao\inst{2} \and Jie Chen\inst{3}}
\authorrunning{Hui Liu et al.}
\institute{College of Electronic and Information, Foshan Polytechnic, \\
    Foshan 528137, P.R. China\\
  \email{lxyliuhui@163.com, \{cycaoyi, lxyliuhui\}@hotmail.com}
  \and
  Department of Computer and Information Sciences, Northumbria University, \\
    Newcastle, NE1 8ST, U.K.\\
  \email{hua.mao@northumbria.ac.uk}
  \and
  College of Computer Science, Sichuan University, \\
  Chengdu 610065, P.R. China\\
  \email{chenjie2010@scu.edu.cn}}

\begin{document}

\maketitle

\begin{abstract}
Compiling the statistics of large-scale IP address data is an essential task in network traffic measurement. The statistical results are used to evaluate the potential impact of user behaviors on network traffic. This requires algorithms that are capable of storing and retrieving a high volume of IP addresses within time and memory constraints. In this paper, we present two efficient algorithms for collecting the statistics of large-scale IP addresses that balance time efficiency and memory consumption. The proposed solutions take into account the sparse nature of the statistics of IP addresses while maintaining a dynamic balance among layered memory blocks. There are two layers in the first proposed method, each of which contains a limited number of memory blocks. Each memory block contains 256 elements of size $256 \times 8$  bytes for a 64-bit system. In contrast to built-in hash mapping functions, the proposed solution completely avoids expensive hash collisions while retaining the linear time complexity of hash-based solutions. Moreover, the mechanism dynamically determines the hash index length according to the range of IP addresses, and can balance the time and memory constraints. In addition, we propose an efficient parallel scheme to speed up the collection of statistics. The experimental results on several synthetic datasets show that the proposed method substantially outperforms the baselines with respect to time and memory space efficiency.

\vspace{6pt}\textbf{Keywords:}
large-scale IP addresses, memory blocks, hash table, sorting, network traffic.
\end{abstract}

\section{Introduction}
\label{sec:Introduction}
In recent years, the amount of network traffic has increased significantly because of the rapid development of emerging network services, such as video streaming, instant messaging, and online payment services. It is crucial to evaluate the potential impact of the features of user behavior on network traffic management. User behavior features are usually extracted from IP packets, which contain IP addresses. There is a close correspondence between informative user behavior features and the IP addresses that frequently appear in the packets. Hence, how to effectively obtain the statistics of large-scale IP address data in a timely manner, such as every few minutes, is a challenging problem in network traffic measurement. Obtaining the statistics of large-scale IP address data typically consists of two tasks: counting the number of occurrences of each IP address and sorting the results in a specific order.

Each IP address serves as a unique identifier for a device on a network. Analyzing large-scale IP address data can reveal vulnerabilities in the network traffic measurements; this enables administrators to discover and resolve weak spots before they are exploited. As a result, potential cybersecurity risks can be effectively reduced in various network applications. Moreover, analyzing such data helps in understanding traffic patterns and behaviors in network traffic measurement process. The collection and analysis of large-scale IP address data provide valuable insights that can guide more reliable and efficient network systems; this helps administrators balance loads and improve the overall performance of the network. Therefore, collecting statistics from large-scale IP address data is an essential task for the efficient, secure, and scalable management of networks.

A number of statistical algorithms for solving this problem have been studied in the last few decades \cite{Jing05DLBS}, \cite{Kapur2013SA}, \cite{Klein2013Sorting}, \cite{Fredman2014Sorting}, \cite{Agapitos2016RSA}. A classic divide-and-conquer strategy has been proposed in which IP addresses are first divided into multiple subsets. Then, each subset is individually computed using a statistics collection method, such as a top $k$ trie algorithm \cite{Jing2013SEDS}. Finally, to merge and sort the results of the multiple subsets, sorting algorithms (e.g., bubble sort, insertion sort, merge sort, selection sort, or quick sort) are used \cite{Hammad2015VS}, \cite{Cormen2001IA}. However, the reduction of computational cost is an intractable problem in the sorting procedure \cite{Puglisi2005SSA}, \cite{Louza2017Sorting}, \cite{Abdel2017CFSA}. For example, the average complexity and worst-case complexity of the bubble, insertion, and selection sort algorithms are $O \left( {{n^2}} \right)$ \cite{Hammad2015VS}, \cite{Shabaz2019SA}, where $n$ represents the number of unsorted records. This indicates that the merging and sorting step of the multiple subsets occupies a large amount of memory and is extremely computationally costly to perform when collecting the statistics of millions or tens of millions of IP addresses. Therefore, statistics collection algorithms using the divide-and-conquer strategy still face several challenges caused by the rapid increase of the large-scale records, such as bounded memory and computational cost restrictions.

A hash table is an effective method for collecting the statistics of IP addresses \cite{Sanders2015HS}. It uses a hash function to compute a hash codes for an array of buckets with the statistical results. The hash function assigns each key to a unique bucket for each IP address. Unfortunately, the hash function can generate the same hash code for more than one IP address. With the increase in the generation of big data, millions or tens of millions of records have become ubiquitous in network traffic. Therefore, this approach could cause several hash collisions, especially for a large number of IP addresses. Although many strategies can be employed to avoid collisions, such as linear probing, quadratic probing, and double hashing, they require extra storage space and computation.

The statistics collection algorithm should be stable, effective, and efficient for large-scale records. Recent advancements in sorting techniques have concentrated on improving the efficiency and scalability of algorithms for processing large-scale data.To overcome the disadvantages of general statistics collection methods, a number of parallel techniques have been developed for large-scale records by optimizing the efficiency and complexity. For example, these algorithms have been extended to the corresponding parallel structures for parallel hardware architectures, such as many-core and multi-core platforms \cite{Alhabboub2024LS}, \cite{Yang2024SORT}, \cite{Satish2009SA},\cite{Singh2018GPU}.

An IP address consists of a series of numbers separated by periods. However, the sorting techniques ignore the importance of the original characteristics of the unsorted IP records. Additionally, parallel statistics collection algorithms are simply parallel computing implementations of the original algorithms. The uniqueness of the IP addresses is vital for distinguishing between the different devices. An IPv4 address is typically represented in four parts. Consequently, collecting statistics from large-scale IP address data still face two significant challenges. First, the unique structure of unsorted IP records is ignored when large-scale IP address data are collected. Second,  the extension of the collection task to the corresponding parallel computation paradigm deserves a further investigation into parallel hardware architectures.

In this paper, we present two efficient algorithms for collecting the statistics of large-scale IP address data. We can obtain the frequently occurring IP addresses from the statistics, which can be regarded as a pre-processing step of user behavior analysis in network traffic management. Because of the increasing volume and speed of network traffic, it has become expensive and impractical to handle all IP addresses contained the IP packets. By taking full advantage of the successive characteristics of memory addresses and the fixed range of each individual part of an IP address, we design two relationship mapping mechanisms between memory blocks and IP addresses for a four-dimensional sparse matrix. The sparse matrix stores the number of occurrences of the individual IP addresses, in which the positions of the rows and columns are employed to represent the mapping relationship between the memory blocks and IP addresses. Specifically, we construct a two-layer memory block (TLMB) to implement the first mapping mechanism for the IP addresses. In addition, we employ a single shared memory block (SSMB) for all IP addresses to implement the other mapping mechanism of the IP addresses. The mechanisms of the mapping relationship effectively remove the information about trivial user behaviors that are irrelevant for statistical analysis. The proposed methods can be extended to the corresponding parallel versions for specific hardware architectures. Extensive experiments on several synthetic datasets show the effectiveness of the proposed method.

Our contributions are summarized as follows.
\begin{enumerate}

\item We present two efficient methods for collecting the statistics of large-scale IP address data that use two different relationship mapping mechanisms, TLMB and SSMB, between memory blocks and IP addresses for a four-dimensional sparse matrix.

\item The computational cost of TLMB is linearly proportional to the number of IP addresses with a limited memory resource, and the memory use of SSMB remains almost unchanged with a reasonable computational cost, regardless of the number of IP addresses.

\item A parallel computation optimization scheme for multiple computers is proposed to effectively improve the computational efficiency and dramatically reduce memory use.

\item Our extensive experimental results using synthetic datasets demonstrate that our proposed method shows clear superior performance in comparison with the baselines, striking a balance between computational cost and memory use.

\end{enumerate}

The remainder of this paper is organized as follows. We briefly describe some of the original sorting techniques as well as various parallel sorting algorithms in Section \ref{sec:work}. In Section \ref{sec:method}, we introduce the proposed method. The experimental results are presented in Section \ref{sec:Experiments}. Finally, we draw the conclusions of the study in Section \ref{sec:Conclusions}.

\section{Related work}
\label{sec:work}

\subsection{Classical sort techniques}
Bubble sort is a classical sorting algorithm in which each element in a list is compared with its neighboring elements and swapped until they are in the desired order \cite{Shutler2008SA}. Bubble sort leads to $\left( {n - 1} \right)$ number of passes and $\frac{{n\left( {n - 1} \right)}}{2}$ number of iterations if $n$ elements are given. Insertion sort is a simple and efficient sorting algorithm that iteratively takes one element and finds its appropriate position in the sorted list by comparing it with neighboring elements. It becomes less efficient as the number of records increases. Selection sort determined the smallest number in an unsorted list and swaps it with the first number in the sorted list. Then, it finds the next smallest element from the remaining list and swaps with the second element in the sorted list. Consequently, the number of sorted elements at the top of the list increases while the rest remain unsorted. Merge sort, which is based on the divide-and-conquer principle, repeatedly divides the array into two halves and then combines them in a sorted manner For more details of classical sort algorithms, such as quick sort and heap sort, we refer the reader to the comments in \cite{Idrizi2017SA}, \cite{Cormen2001IA}. It is well known that computational cost and required memory are the primary concerns in sorting algorithms \cite{Rusu2018Sorting}, \cite{Jukna2020Sorting}. Table \ref{tb:summary} shows the time complexity of the best case, average case, and worst case of several classical sorting algorithms \cite{Cormen2001IA}, \cite{Kocher2014SA}, \cite{Faujdar2015SA}. These sorting algorithms can be used to find the first $k$ IP addresses of the most frequent occurrences from the statistical results obtained for IP address data.

\begin{table}[!t]
\small
\setlength{\abovecaptionskip}{2pt}
\setlength{\belowcaptionskip}{10pt}
\setlength{\tabcolsep}{2pt}
\centering
\caption{Time complexity of various sorting algorithms \cite{Cormen2001IA}, \cite{Kocher2014SA}, \cite{Faujdar2015SA}.}
\label{tb:summary}
\begin{tabular}{ccccc}
\hline
Algorithm & $Best \left( n \right)$ & $Average \left( n \right)$ & $Worst \left( n \right)$ \\
\hline
Bubble sort  & $O \left( {n} \right)$ & $O \left( {n^2} \right)$ & $O \left( {n^2} \right)$ \\
Insertion sort  & $O \left( {n} \right)$ & $O \left( {n^2} \right)$ & $O \left( {n^2} \right)$ \\
Selection sort & $O \left( {n^2} \right)$ & $O \left( {n^2} \right)$ & $O \left( {n^2} \right)$ \\
Merge Sort & $O\left( {n\log n} \right)$ & $O\left( {n\log n} \right)$  & $O\left( {n\log n} \right)$  \\
Quick Sort & $O \left( {n\log n} \right)$ & $O\left( {n\log n} \right)$  & $O\left( {n^2} \right)$  \\
Heap Sort & $O\left( {n\log n} \right)$ & $O\left( {n\log n} \right)$ & $O\left( {n\log n} \right)$ \\
\hline
\end{tabular}
\end{table}

\subsection{Accelerating large-scale sorting techniques}
Sorting is an essential part of modern computing. Significant efforts have recently been dedicated to accelerating large-scale sorting techniques \cite{Alhabboub2024LS}, \cite{Jug2024ASS},  \cite{Yang2024SORT}. For example, Alhabboub \textit{et al.} improve the computation efficiency of  the classical QuickSort algorithm by combining with parallel implementations \cite{Alhabboub2024LS}. The improved QuickSort algorithm can be applied on the sorting  large-scale data, and exhibits slightly superior computational efficiency compared to classical sequential QuickSort. Jug\'{e} \textit{et al.} proposed an adaptive ShiversSort algorithm for efficiently sorting partially sorted data, which is considered as a variant of the well-known algorithm TimSort \cite{Jug2024ASS}. Yang \textit{et al.} proposed a high-performance parallel sorting algorithm on a CPU-DSP heterogeneous processor \cite{Yang2024SORT}. These methods typically take into account different sorting settings, such as parallel sorting environments, data criterions, or specific CPU architectures. These large-scale sorting algorithms provide an efficient post-processing step for collecting the statistics of large-scale IP address data.

\subsection{Parallel hardware architectures}
The hardware architecture of modern processors usually consists of more than two independent central processing units (CPUs) or graphics processing units (GPUs). Parallel software platforms can be implemented using high-level programming frameworks for specific hardware architectures \cite{Chen2009SA}. The Compute Unified Device Architecture (CUDA) is a parallel computing platform for general computing on GPUs. Most parallel sorting algorithms are variants of standard, well-known sorting algorithms adapted to GPU hardware architecture. For example, Cederman designed a quick sort for the GPU platform \cite{Cederman2008}, and Peters proposed an adaptive bitonic sorting algorithm with a bitonic tree for GPUs \cite{Peters2011BS}. The parallel computation of sorting algorithms is considered to be the most efficient way of sorting elements on parallel hardware architectures \cite{Singh2018GPU}.

\begin{figure}[!htbp]
\centering
\includegraphics[width=0.2\textwidth]{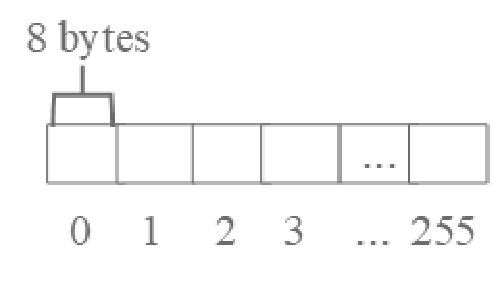}
\caption{Example of a memory block of size $256 \times 8$ bytes containing 256 elements.}
\label{fig:memoryblock}
\end{figure}

\section{Proposed method}
\label{sec:method}

\subsection{Problem Formulation}
IP flow data $FD$ is a sequence of IP records, that is, $FD = \left\{ \left( {x_1, p_1} \right), ..., \left( {x_n, p_n} \right) \right\}$ and ${n \ge 1{e^6}}$, where each pair of elements $\left( {x_i, p_i} \right)$ $\left( {i \in \left[ {1,n} \right]} \right)$ consists of an IP address $x_i$ and a set of corresponding user behavior attributes $p_i$. Given a finite set of IP addresses $X = \left\{ {{x_1},{x_2},...,{x_n}} \right\} \in {\mathbb{R}^{m \times n}}$, the purpose of the IP address statistics task is to efficiently determine the first $k$ IP addresses of the most frequent occurrences in $X$, where $m$ the dimensionality of an individual IP address and $k \ll n$.

\subsection{IP Address Statistics Task}
A standard IP address is composed of four decimal numbers ranging from 0 to 255 which are separated by dot symbols. An individual IP address is logically divided into four parts by splitting it with respect to each dot symbol, and each part of an IP address has an integer value. We create a four-dimensional array for the statistics of IP addresses, where the length of each dimension in the array is 256. Each element of the array can be employed to store the number of occurrences of the IP address according to the relationship mapping between the index of each dimension of the array and the integer value of the corresponding part of the IP address. For example, consider the individual IP address 1.2.3.4 and the four-dimensional array $fd\_{array}$. The number of occurrences of this IP address is stored in ${fd}\_{array}[1][2][3][4]$. However, individual IP addresses in the host logs often make up a small proportion of all IP addresses. The array can be considered to be sparse because most of its elements are zeros. Consequently, we can carefully design a four-dimensional sparse matrix to store the number of an individual IP address by taking full advantage of the successive characteristics of array addresses and the fixed range of an individual part of an IP address.

\begin{algorithm}[!htbp]
\renewcommand{\algorithmicrequire}{\textbf{Input:}}
\renewcommand\algorithmicensure {\textbf{Output:} }
\caption{TLMB}
\label{alg:ESS}
\begin{algorithmic}[1]
\REQUIRE ~~\\
A finite set of IP addresses $X = [{x_1},{x_2},...,{x_n}] \in {\mathbb{R}^{m \times n}}$, number $k > 1$.
\end{algorithmic}
\begin{algorithmic}[1]
\STATE Construct $256 \times 256$ memory blocks of size 128 MB for the first layer;
\FOR { $i = 1 : n$ }
\STATE Assume that $a$, $b$, $c$ and $d$ each represent one of the integer values of the four parts of IP address $x_i$.
\STATE Calculate the index of the memory block in the first layer: $p = a \times 255 \times 255 + b \times 255$.
\IF {the value of the $j$-th element is null in the $p$-th memory block}
\STATE Create a memory block in the second layer, set all elements of the memory block to zero, and store the starting address of the memory block in the $j$-th element.
\ELSE
\STATE Obtain the starting address of the memory block $m$ in the second layer by finding the $j$-th element of the $p$-th memory block.
\ENDIF
\STATE Add 1 to the $d$-th element of memory block $m$ in the second layer.
\ENDFOR
\STATE Construct a minimum heap of size $k$ using each non-zero element of the memory blocks in the second layer.
\ENSURE ~~\\
\STATE Traverse the nodes of the heap to obtain the $k$ IP addresses and their number of occurrences.
\end{algorithmic}
\end{algorithm}

\subsection{IP Usage Storage and Retrieval Strategies for the Four-dimensional Sparse Matrix}
\label{sec:ius}
Assume that each memory address of a 64-bit system can be stored in an element 8 bytes in size, and the number of occurrences of an individual IP address is no more than $2^{64}$. There is an array of size 256 elements that consists of $256 \times 8$ bytes of memory. The array is regarded as a memory block that contains a contiguous address space, as shown in Fig. \ref{fig:memoryblock}. In other words, the addresses of all bytes of the array are sequential in the memory block. Therefore, the position of the array can be indexed by the integer value of a particular part of the IP address. We present two efficient methods to collect the statistics of large-scale IP address data, each of which contains a relationship mapping mechanism between memory blocks and IP addresses for the four-dimensional sparse matrix.

\begin{figure*}[!htbp]
\centering
\includegraphics[width=1.0\textwidth]{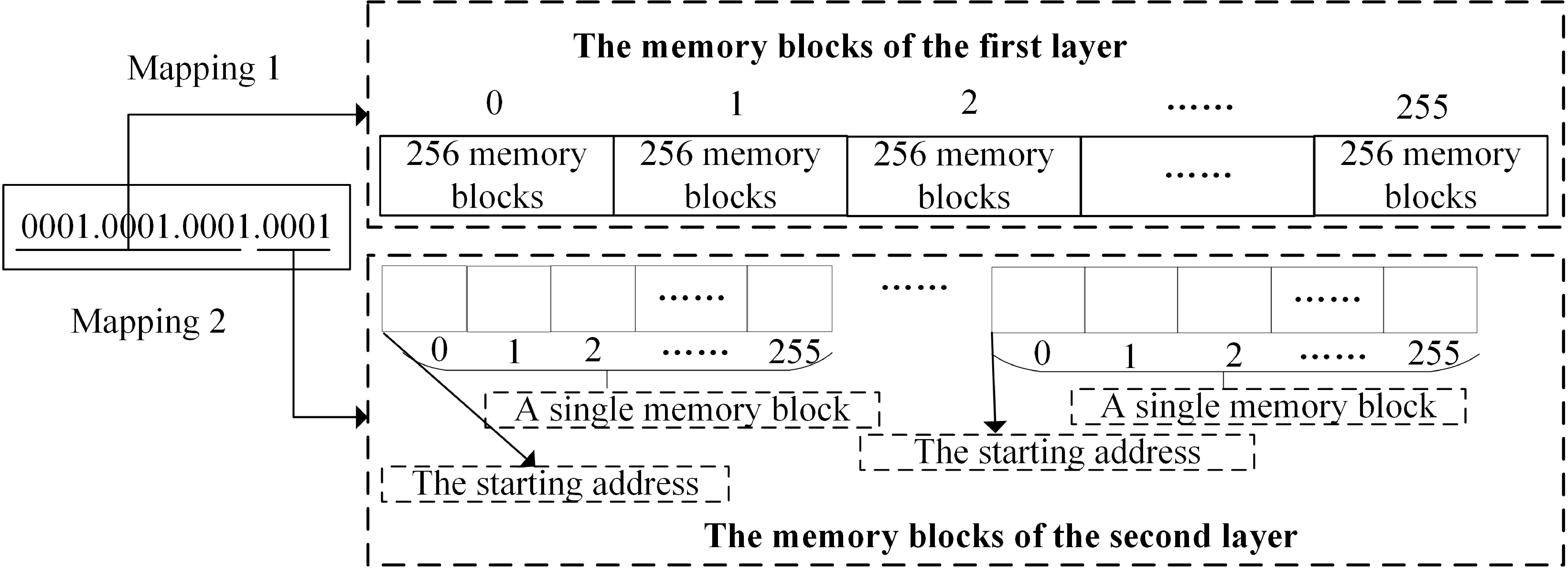}
\caption{An example of the mapping relationships between memory blocks and an IP address.}
\label{fig:architecture}
\end{figure*}

\subsubsection{First method: TLMB}
The first proposed mapping mechanism of IP addresses is TLMB. The four parts of the IP address are represented in four layers, where each layer is made up of one or more memory blocks. The first layer only contains one memory block, whereas the second layer contains 256 memory blocks. Each memory block contains 256 elements. Each element of the memory block in the first layer is employed to store the starting addresses of the corresponding 256 memory blocks in the second layer. Similarly, the third layer contains $256 \times 256$ memory blocks, the size of which is 128 MB in memory. Then, the element of each memory block in the third layer stores the starting address of the corresponding memory block in the fourth layer. This would be 32 GB in size if we adopted a pre-allocation strategy for all memory blocks in the four layers. Hence, we present an alternative pre-allocation strategy for the memory blocks. A memory block will be allocated only when the first three parts of an initial IP address have been given. In particular, pre-allocating a big memory block of size 128 MB containing $256 \times 256$  contiguous memory blocks is feasible in a modern computer. Consequently, the first two layers can be removed from this architecture if the third layer has contiguous memory blocks of 128 MB.

We formally present a storage strategy for IP addresses that consists of two layers that consist of a limited number of memory blocks. The first layer contains $256 \times 256$ memory blocks. The first three parts of the IP address can be mapped into the corresponding position of the element in a particular memory block of the first layer according to the individual values of the three parts. We allocate a memory block in the other layer for the IP address when its first three parts are initially given. Each element of a memory block in this layer stores the number of occurrences of the corresponding IP address. Figure \ref{fig:architecture} shows an example of the relationship mapping between the memory blocks of two layers and an IP addresses.

Consider the individual IP address 1.2.3.4, we have $1 \times 255 \times 255 + 2 \times 255 = 65535$, which represents the index of the memory block in the first layer. The positions of the first two dimensions of the sparse matrix can be mapped to the elements of the memory blocks included in the first layer. The third part of the IP address denotes the index of the memory block in the second layer. The positions of the final dimensionality of the sparse matrix can be indexed by combining the starting address of the memory block in the second layer with the integer value of the four parts of the IP address.

\begin{algorithm}[!htbp]
\renewcommand{\algorithmicrequire}{\textbf{Input:}}
\renewcommand\algorithmicensure {\textbf{Output:} }
\caption{SSMB}
\label{alg:ESS2}
\begin{algorithmic}[1]
\REQUIRE ~~\\
A finite set of IP addresses $X = [{x_1},{x_2},...,{x_n}] \in {\mathbb{R}^{m \times n}}$, number $k > 1$.
\end{algorithmic}
\begin{algorithmic}[1]
\STATE Construct a memory block of size 128 MB saved by all IP addresses;
\STATE All IP addresses are logically partitioned into $q$ subsets according to the first part of each IP address.
\FOR { $i = 1 : n$ }
\STATE Assume that $a$, $b$, $c$ and $d$ each represent one of the integer values of the four parts of IP address $x_i$.
\FOR { $j = 1 : q$ }
\IF {$j == q $}
\STATE Calculate the position of the memory block in the first layer: $p = b \times 255 \times 255 + c \times 255 + d$.
\STATE  Add 1 to the $p$-th element of the memory block.
\ENDIF
\ENDFOR
\STATE Construct a minimum heap of size $k$ using the each non-zero elements of the memory block.
\ENDFOR
\ENSURE ~~\\
\STATE Traverse of the nodes of the heap to obtain the $k$ IP addresses and their number of occurrences.
\end{algorithmic}
\end{algorithm}

We traverse all elements of the memory blocks of the second layer to obtain the maximum number of occurrences of elements if $k=1$. Otherwise, we construct a minimum heap of size $k$. The statistical results of the first $k$ IP addresses are saved in the heap, which is a special binary tree and implemented by an array of size $k$. The construction of the heap is completed by traversing all elements of the memory blocks of the second layer. The tree node in the leap contains two important attributes: an IP address and its number of occurrences. The final IP addresses and numbers of occurrences can be obtained by a traversal of the nodes of the heap. The complete procedure for determining the first $k$ of the most frequent IP addresses from host logs is outlined in Algorithm \ref{alg:ESS}.

\subsubsection{Second method: SSMB}
We also designed an SSMB that stores all IP addresses and their statistics. IP addresses are logically divided into at most 256 subsets according to the value of the first part of each individual IP address. We construct a single memory block of $256 \times 256 \times 256 $ elements that is $256 \times 256 \times 256 \times 8$  bytes in size, that is, 128 MB. The memory block is shared by all subsets. For each subset, the last three parts of the IP address can be mapped into the corresponding position of the element in the shared single memory block, which is always initialized at the beginning of the relationship mapping. Then, the element of this memory block stores the statistics of the IP addresses in this subset.

We further perform a round traversal of the memory block to initialize a minimum heap of size $k$ after the relationship mapping has been completed in the first subset. Then, we continue to perform a round traversal of the memory block to adjust the heap after the relationship mapping has been completed for the subsequent subsets. Finally, we obtain the first $k$ most frequently occurring IP addresses in the heap. The complete procedure for finding the first $k$ of the most frequency IP addresses from host logs is outlined in Algorithm \ref{alg:ESS2}.

\subsection{Memory Use and Complexity Analysis}
We first evaluate the memory use and computational complexity of the first proposed method, TLMB. The size of each memory block is $256 \times 8$ bytes, and there are $256 \times 256$ memory blocks in the first layer. Hence, the size of the memory blocks in the first layer is 128 MB. Assume that the number of the distinct first three parts of the IP addresses is $s$. The number of memory blocks in the second layer is linearly proportional to $s$. Moreover, the memory size of the minimum heap is $\left( k + 8 \right)$ bytes, where each tree node contains two attributes, that is, an IP address and the number of occurrences. The total size of the memory of the proposed method is approximately the sum of the three parts, that is, 128 MB, $s$ KB, and $\left( k + 8 \right)$ bytes. The computational complexity of the two layers for calculating the IP address statistics is ${O}\left( n \right)$ in Algorithm \ref{alg:ESS}, where $n$ is the number of IP addresses. In addition, the computational complexity of constructing a minimum heap of size $k$ is ${O}\left( {k\log k} \right)$, where $k$ is the number of tree nodes in the heap. Consequently, the overall computational complexity of the proposed algorithm is ${O}\left( {k\log k + n} \right)$.

We next evaluate the memory use and computational complexity of the second proposed method, SSMB. The memory size of SSMB is 128 MB. Assume that the number of the distinct first parts of the IP addresses is $q$. The computational complexity of the mapping mechanism of the IP addresses in Algorithm \ref{alg:ESS} is ${O}\left( qn \right)$, where $n$ is the number of IP addresses. Similarly, the computational complexity of constructing a minimum heap of size $k$ is ${O}\left( {k\log k} \right)$ for each subset. Consequently, the overall computational complexity of the proposed algorithm is ${O}\left( { q \left( k\log k +  n \right)} \right)$.

\subsection{Parallel computation optimization techniques}
\subsubsection{Parallel computation mechanism of TLMB}
In the worst case, the distinct first three parts of the IP addresses cover all the binary combinations. Hence, the size of the memory blocks in the second layer is 32 GB. We present a parallel computation scheme on multiple computers for improving the computational efficiency and reducing memory use. Assume that there are $2^r$ computers available for parallel computation, where $r$ represents a positive integer. The task of collecting IP address statistics is then divided into multiple subtasks, which are performed by $2^r$ computers, respectively, according to the first $r$ bits of the first part of the IP addresses. The number of memory blocks reduces to $\left(256 \times 256 \right) / 2^r$ for $2^r$ computers. Simultaneously, the number of memory blocks will decrease to $32 / 2^r$ GB in the second layer. For example, the number of memory blocks in the first layer is $256 \times 64$ for each computer when $r = 2$, and the size of memory blocks in the second layer is 8 GB in the worst case. In addition, if four computers perform the task of computing IP address statistics by partitioning the first two bits of the first part of the IP addresses, then the second layers of multiple computers are merged into a complete second layer, where is employed to construct a minimum heap of size $k$. Finally, the parallel computation results can be obtained in a manner similar to the last step of Algorithm \ref{alg:ESS}.

\subsubsection{Parallel computation mechanism of SSMB}
Assume that all IP addresses are logically divided into $q$ subsets according to the value of the first part of an individual IP address. Further assume are $q$ computers for parallel computation, where the statistics collection task of each subset can be performed by an individual computer. A minimum heap of size $k$ is shared among these computers. Hence, this greatly increases the computational efficiency of the task by $q$ times.

\section{Experiments}
\label{sec:Experiments}

\subsection{Experimental Settings}
In this section, we evaluate the performance of the proposed methods \footnote{https://github.com/chenjie20/IPStatistics} on two different types of datasets, i.e., three synthetic datasets and two real-world datasets. The three synthetic datasets contain 5 million, 10 million, and 50 million randomly generated IP records. Each individual IP address contains one or more of IP records. The average number of IP records is 100 for each individual IP address. In particular, the IPv4 addresses were specifically divided into four segments in the experiments. These experimental settings ensure a comprehensive evaluation of the capacity of TLMB and SSMB to efficiently collect statistics for large-scale IP address data. Parameter $k$ represents the number of frequently occurring IP addresses. Additionally, two real-world network traffic datasets,  provided by the Center for Applied Internet Data Analysis (CAIDA) \footnote{https://www.caida.org/catalog/datasets/ipv4\_prefix\_probing\_dataset}, are collected from various parts of the internet. These two datasets are widely used in networking and traffic analysis research. The statistics of these datasets are summarized in Table \ref{tb:statistics}.

\begin{table}[!t]
\small
\setlength{\abovecaptionskip}{5pt}
\setlength{\belowcaptionskip}{5pt}
\setlength{\tabcolsep}{3pt}
\centering
\caption{Statistics of the datasets.}
\label{tb:statistics}
\begin{tabular}{cccccc}
\hline
Data sets & IP Records & Individual IP Addresses & Size & Type \\
\hline
1 & 5, 000, 000 & 50, 000 & 77.5 MB & Synthetic \\
2 & 10, 000, 000 & 100, 000 & 155 MB & Synthetic \\
3 & 50, 000, 000 & 500, 000 & 775 MB & Synthetic \\
\hline
4  & 1, 114, 633 & 107, 988 & 14.4 MB & Real \\
5  & 1, 430, 258 & 133, 116 & 18.5 MB & Real \\
\hline
\end{tabular}
\end{table}

We compared the proposed method with the following baselines:
\begin{enumerate}[$\bullet$]

\item \textbf{Hash Mapping}. Each IP record is mapped into an entry with a statistical result using a hash table \footnote{https://github.com/activesys/libcstl}. Next, the statistical results are used to construct a minimum heap of size $k$.

\item \textbf{IP Mapping}. All IP records are partitioned into $q$ subsets according to the first part of each IP address. The statistics of the IP records in each subset are mapped into an array, whose memory is pre-allocated on a computer according to the last three parts of each IP address. The first $k$ most frequent IP addresses are chosen from each subset. Next, a minimum heap of size $k$ is constructed using the $q\times k$ IP addresses.

\end{enumerate}

Memory blocks were pre-allocated for TLMB and SSMB, with sizes ranging from small-scale (e.g., 5 million) to large-scale (e.g., 50 million) to test scalability. Two metrics were employed to evaluate the sorting performance, that is, computational cost and memory use. All experiments were implemented using the C language on a Windows platform with an Intel i7-9700k CPU and 32 GB RAM.

\begin{table*}[!t]
\small
\setlength{\abovecaptionskip}{5pt}
\setlength{\belowcaptionskip}{5pt}
\setlength{\tabcolsep}{3pt}
\centering
\caption{Computational cost (s) of different methods on the three synthetic datasets.}
\label{tb:results:cost}
\begin{tabular}{cccccc}
\hline
Data & $k$  & Hash Mapping & IP Mapping & Ours (TLMB) & Ours (SSMB) \\
\hline
\multirow{2}{*}{1} & 10 & 1,458.14 (2.32) & \underline{15.18} (0.07) & \textbf{2.51} (0.01) & 18.43 (0.05) \\
~ & 100 & 1,458.56 (2.76) & \underline{15.21} (0.04) & \textbf{2.52} (0.01) & 18.43 (0.06) \\
\hline
\multirow{2}{*}{2} & 10 & 2,927.23 (11.64) & \underline{17.41} (0.11) & \textbf{4.92} (0.01) &  27.59 (0.05) \\
~ & 100  & 2,934.65 (28.93) & \underline{17.45} (0.05) & \textbf{4.95} (0.01) & 27.65 (0.06) \\
\hline
\multirow{2}{*}{3} & 10 & 14,547.09 (13.95) & \underline{35.33} (0.09) & \textbf{24.22} (0.07) &  101.01 (0.17) \\
~ & 100 & 14,566.36 (26.78) & \underline{35.39} (0.05) & \textbf{24.24} (0.04) & 101.20 (0.2) \\
\hline
\end{tabular}
\end{table*}

\begin{table*}[!t]
\small
\setlength{\abovecaptionskip}{5pt}
\setlength{\belowcaptionskip}{5pt}
\setlength{\tabcolsep}{3pt}
\centering
\caption{Memory use (MB) of different methods on the three synthetic datasets.}
\label{tb:results:memory}
\begin{tabular}{cccccc}
\hline
Data & $k$& Hash Mapping & IP Mapping &  Ours (TLMB) & Ours (SSMB) \\
\hline
\multirow{2}{*}{1} & 10 & \textbf{43.63} (0.12) & 16,332.81 (0.07) & 189.61 (0.16) &  \underline{139.77} (0.13) \\
~ & 100 & \textbf{43.73} (0.06) & 16,332.77 (0.05) & 189.64 (0.07) &  \underline{139.79} (0.07) \\
\hline
\multirow{2}{*}{2} & 10 & \textbf{74.8} (0.12) & 16,332.75 (0.12) &  239.11 (0.1) &  \underline{139.74} (0.13) \\
~ & 100 & \textbf{74.83} (0.05) & 16,332.77 (0.07) & 239.06 (0.15) & \underline{139.7} (0.12) \\
\hline
\multirow{2}{*}{3} & 10 & \textbf{324.45} (0.99) & 16,332.59 (0.06) & 628.33 (0.07) &  \underline{139.6} (0.15) \\
~ & 100 & \textbf{324.4} (1.02) & 16,332.6 (0.16) & 628.27 (0.01) &  \underline{139.71} (0.16) \\
\hline
\end{tabular}
\end{table*}

\subsection{Experiment Results}

\subsubsection{Experimental Evaluation on Synthesized Data}
Parameter $k$ was set to 10 or 100, and we repeated each experiment 10 times. The average computational costs and standard deviations are reported in Table \ref{tb:results:cost}, and the mean memory use and standard deviations are given in Table \ref{tb:results:memory}. The results show that TLMB consistently outperformed all the other methods in terms of computational cost. For example, TLMB achieves computational costs of 2.51 s and 2.52 s when $k=10$ and $k=100$, respectively. When the number of IP addresses increases from 5 million to 50 million with $k=10$, the gap in computational costs of TLMB and IP Mapping are 12.6 s and 11.11 s, respectively. We also observed the same advantages when $k=100$. In addition, SSMB shows competitive results when compared with the comparison methods in terms of memory use. The memory use of SSMB is always approximately 139 MB, even when the number of IP addresses changes. In contrast, the computational cost of SSMB substantially outperforms that of Hash Mapping under different numbers of IP addresses. IP Mapping obtained the lowest computation cost for all numbers of IP addresses. However, the highest memory use results of IP Mapping are consistent with expectations.

\begin{table*}[!t]
\small
\setlength{\abovecaptionskip}{5pt}
\setlength{\belowcaptionskip}{5pt}
\setlength{\tabcolsep}{3pt}
\centering
\caption{Computational cost (s) of different methods on the two real-world datasets.}
\label{tb:results:real:cost}
\begin{tabular}{cccccc}
\hline
Data & $k$  & Hash Mapping & IP Mapping & Ours (TLMB) & Ours (SSMB) \\
\hline
\multirow{2}{*}{4} & 10 & 1675.3 & \underline{6.73} & \textbf{0.61}  & 7.00 \\
~ & 100 & 1673.9 & \underline{6.90} & \textbf{0.62}  & 7.09 \\
\hline
\multirow{2}{*}{5} & 10 & 1996.8 & 7.33 & \textbf{0.78}  &\underline{ 6.88} \\
~ & 100  & 1989.9 & 7.47 & \textbf{0.80}  & \underline{7.02} \\
\hline
\end{tabular}
\end{table*}

\begin{table*}[!t]
\small
\setlength{\abovecaptionskip}{5pt}
\setlength{\belowcaptionskip}{5pt}
\setlength{\tabcolsep}{3pt}
\centering
\caption{Memory use (MB) of different methods on the two real-world datasets.}
\label{tb:results:real:memory}
\begin{tabular}{cccccc}
\hline
Data & $k$& Hash Mapping & IP Mapping &  Ours (TLMB) & Ours (SSMB) \\
\hline
\multirow{2}{*}{4} & 10 & \underline{238.9}  & 16,392.1 & 249.8 &  \textbf{183.3} \\
~ & 100 & \underline{238.8}  &  16,392.1 & 249.8 &  \textbf{183.4} \\
\hline
\multirow{2}{*}{5} & 10 & \underline{266.7}  &  16,391.9  & 281.4 &  \textbf{183.4} \\
~ & 100 & \underline{266.7}  & 16,392.3 &  281.4 &  \textbf{183.4} \\
\hline
\end{tabular}
\end{table*}

\subsubsection{Experimental Evaluation on Real-World Data}
We evaluate the proposed and competing methods on two real-world datasets. Tables \ref{tb:results:real:cost} and \ref{tb:results:real:memory} show the computational costs and memory usages levels of different methods, respectively. TLMB consistently incurs lower computational costs than do the other methods. SSMB and IP mapping demonstrate comparable computational costs. However, SSMB significantly reduces the memory requirements relative to IP mapping. Furthermore, Hash mapping incurs a higher computational cost than the competing methods do because of its use of hash computations, making it prohibitively time-consuming in practice. As the number of IP addresses increases across the two real-world datasets, SSMB maintains relatively stable memory usage. These findings highlight the effectiveness of both TLMB and SSMB.

\subsection{Ablation study}
To investigate the impact of the memory blocks in the proposed TLMB and SSMB methods, we performed ablation studies on the three synthetic datasets. Specially, we examined two particular cases in the experiments. The hash table was employed to replace the second part of TLMB and SSMB, respectively. The primary goal of the ablation study is to demonstrate the importance of the memory blocks on collecting the statistics of large-scale IP address data. The invariants of TLMB and SSMB corresponding to these two cases are referred to as TLMB$_{hash}$ and SSMB$_{hash}$, respectively.

Tables \ref{tb:as:cost} and  \ref{tb:as:memory} show the results of the ablation study regarding the computational cost and memory use. TLMB$_{hash}$ exhibits similar computational cost and memory use compared to SSMB$_{hash}$ on the first two synthetic datasets. Additionally, the computational cost of TLMB$_{hash}$ is slightly higher  than that of SSMB$_{hash}$, while its memory usage is marginally lower. TLMB$_{hash}$ and SSMB$_{hash}$ achieve an acceptable computational cost. However, TLMB and SSMB achieves superior performance in computational cost and memory use compared with those of TLMB$_{hash}$ and SSMB$_{hash}$. These results further emphasize that integrating a hash table scheme into TLMB and SSMB is both time-consuming and memory-intensive. Therefore, the results of the ablation study demonstrate the effectiveness of the memory blocks in the proposed TLMB and SSMB methods.

\begin{table*}[!t]
\small
\setlength{\abovecaptionskip}{5pt}
\setlength{\belowcaptionskip}{5pt}
\setlength{\tabcolsep}{3pt}
\centering
\caption{Ablation study concerning the computational costs (s) incurred on the three synthetic datasets.}
\label{tb:as:cost}
\begin{tabular}{cccccc}
\hline
Data & $k$  & TLMB$_{hash}$ & SSMB$_{hash}$ & Ours (TLMB) & Ours (SSMB) \\
\hline
\multirow{2}{*}{1} & 10 & 37.95 & 37.16 & \textbf{2.51} &  \underline{18.43} \\
~ & 100 &  37.9 & 37.34 & \textbf{2.52} &  \underline{18.43} \\
\hline
\multirow{2}{*}{2} & 10 &  75.68 & 74.51  & \textbf{4.92} &  \underline{27.59} \\
~ & 100  &  75.19  &  74.05 & \textbf{4.95} &  \underline{27.65} \\
\hline
\multirow{2}{*}{3} & 10 & 384.72 & 374.49  & \textbf{24.22} & \underline{101.01} \\
~ & 100 &  384.23 &  375.9 & \textbf{24.24} &  \underline{101.20} \\
\hline
\end{tabular}
\end{table*}

\begin{table*}[!t]
\small
\setlength{\abovecaptionskip}{5pt}
\setlength{\belowcaptionskip}{5pt}
\setlength{\tabcolsep}{3pt}
\centering
\caption{Ablation study concerning the memory use (MB) required for the three synthetic datasets.}
\label{tb:as:memory}
\begin{tabular}{cccccc}
\hline
Data & $k$& TLMB$_{hash}$ & SSMB$_{hash}$ &  Ours (TLMB) & Ours (SSMB) \\
\hline
\multirow{2}{*}{1} & 10 & 3,397.3 &  3397.2 &  \underline{189.61} & \textbf{139.77} \\
~ & 100 & 3,397.5  & 3,397.2 &  \underline{189.64} &  \textbf{139.79} \\
\hline
\multirow{2}{*}{2} & 10 & 6,654.4  &  6,654.3 &   \underline{239.11} &  \textbf{139.74} \\
~ & 100 & 6,654.4  &  6,654.4 &  \underline{239.06} & \textbf{139.7} \\
\hline
\multirow{2}{*}{3} & 10 & 25,665.6 & 27,367.6 &  \underline{628.33} &  \textbf{139.6} \\
~ & 100 & 26,008.6 & 26,517.5 & \underline{628.27} &  \textbf{139.71} \\
\hline
\end{tabular}
\end{table*}

\subsection{Empirical investigation}
We empirically examined the effect induced by varying the $k$ considered in the proposed TLMB and SSMB methods. Here $k$ was selected from the set $\left\{ {10,20,50,100,200,500} \right\}$. The computational cost and memory use were employed to evaluate TLMB and SSMB with different $k$ values.

\begin{figure*}[!htbp]
\begin{minipage}[t]{0.49\linewidth}
\centering
\subfigure[]{
\label{fig:cc:a} 
\includegraphics[width=0.8\textwidth]{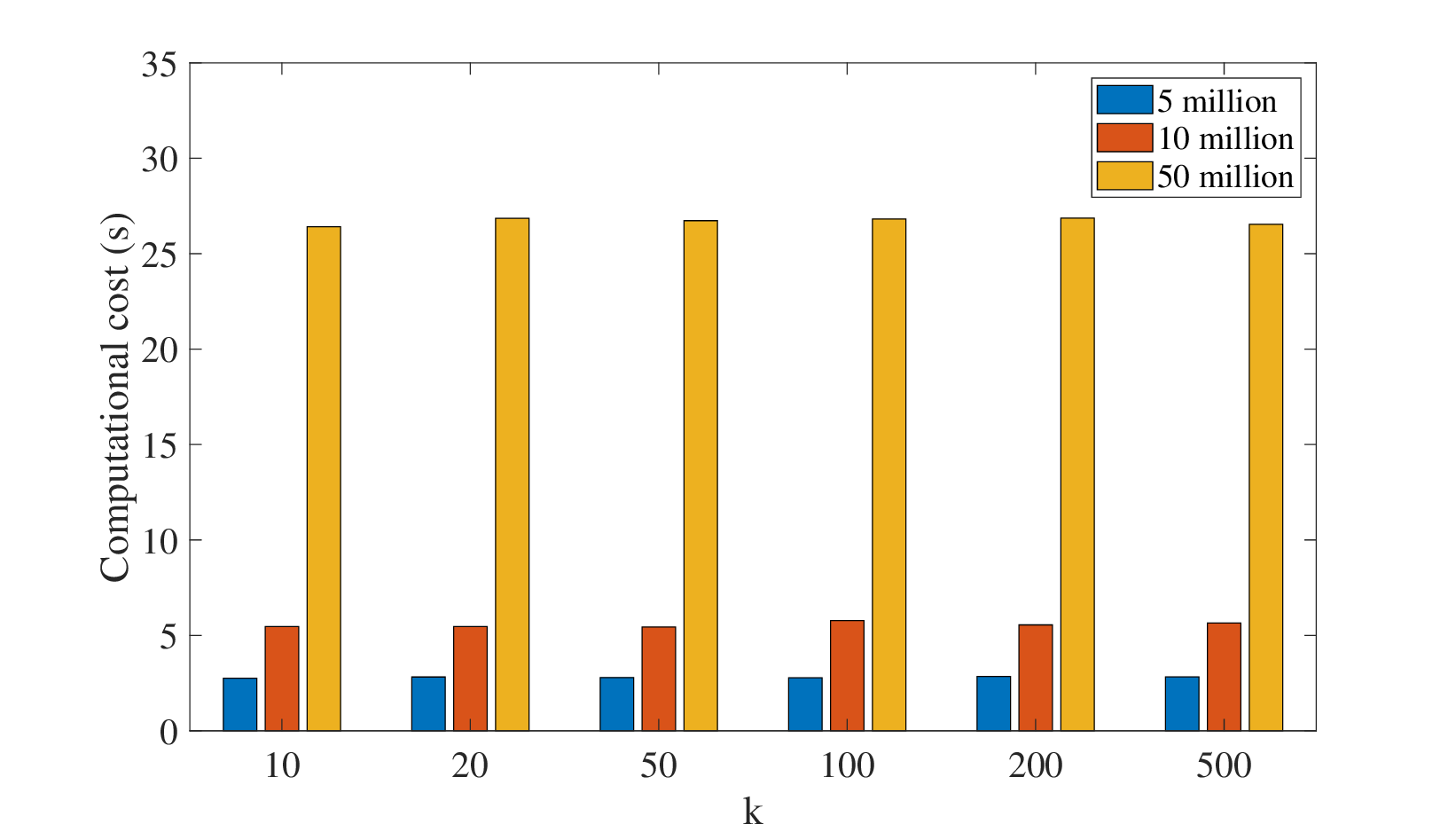}}
\end{minipage}%
\begin{minipage}[t]{0.49\linewidth}
\centering
\subfigure[]{
\label{fig:cc:b} 
\includegraphics[width=0.8\textwidth]{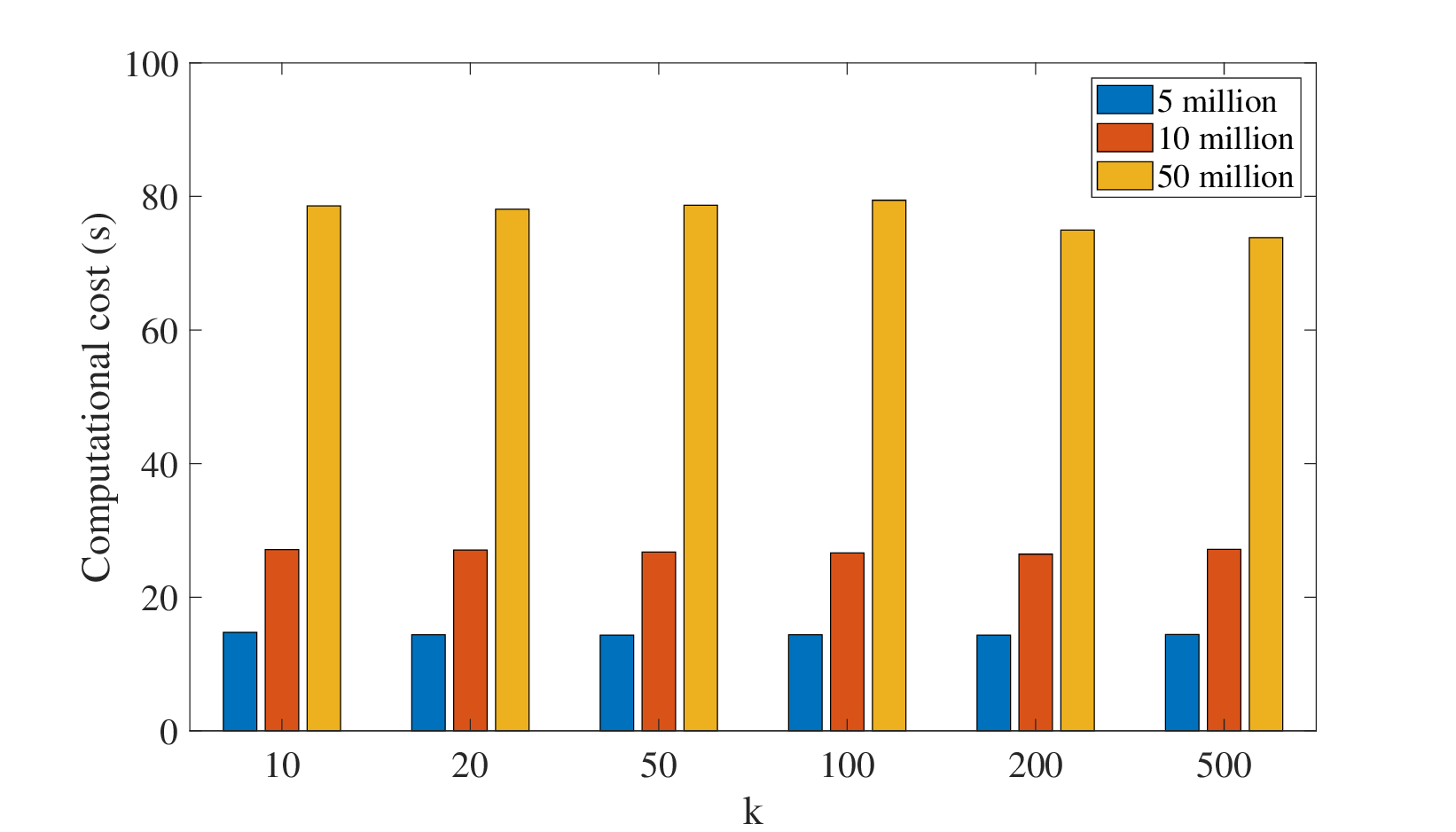}}
\end{minipage}%
\caption{The computational costs of the proposed TLMB and SSMB methods with different $k$ values. (a) TLMB and (b) SSMB.}
\label{fig:cc} 
\end{figure*}

\begin{figure*}[!htbp]
\begin{minipage}[t]{0.49\linewidth}
\centering
\subfigure[]{
\label{fig:mu:a} 
\includegraphics[width=0.8\textwidth]{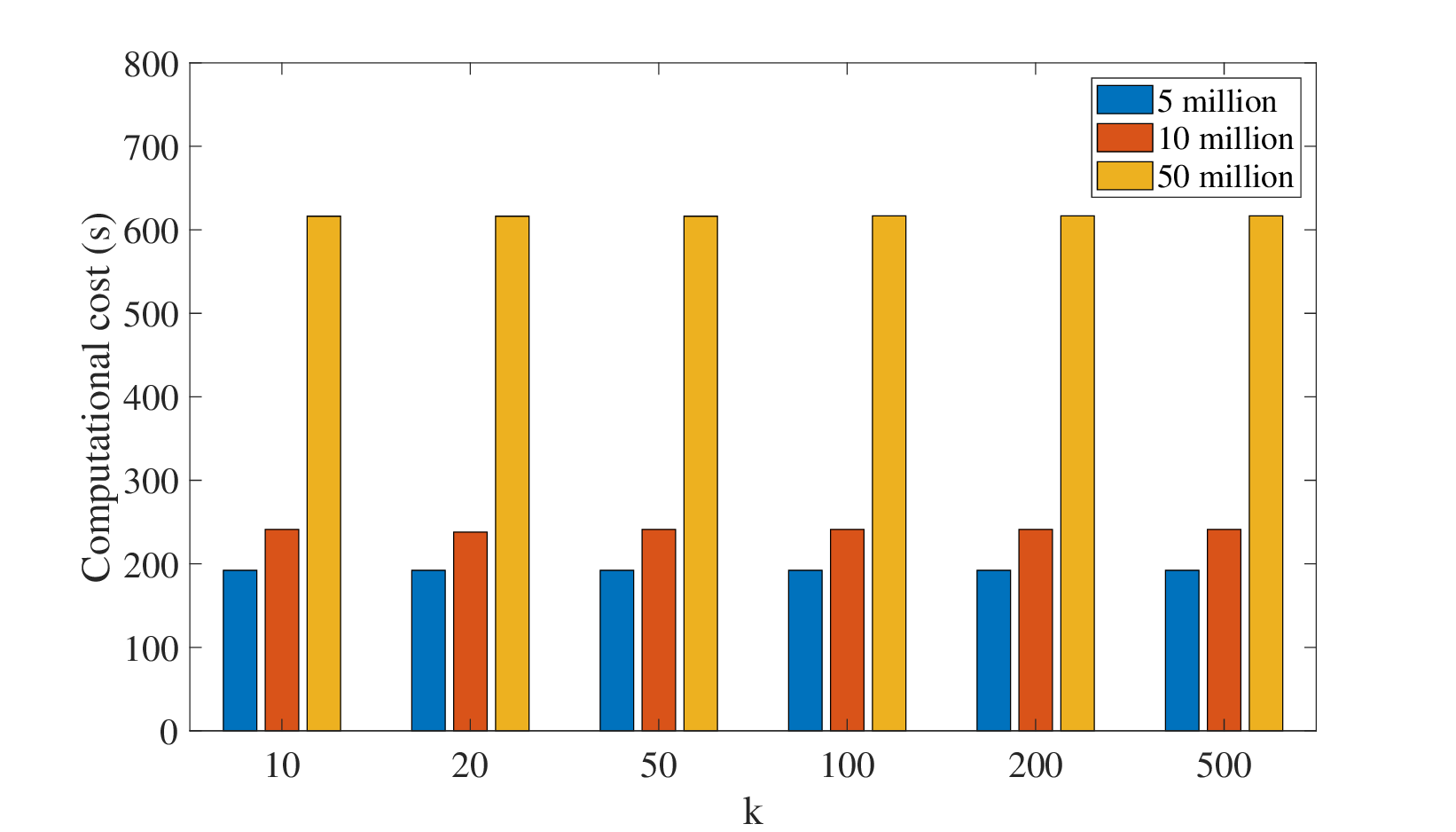}}
\end{minipage}%
\begin{minipage}[t]{0.49\linewidth}
\centering
\subfigure[]{
\label{fig:mu:b} 
\includegraphics[width=0.8\textwidth]{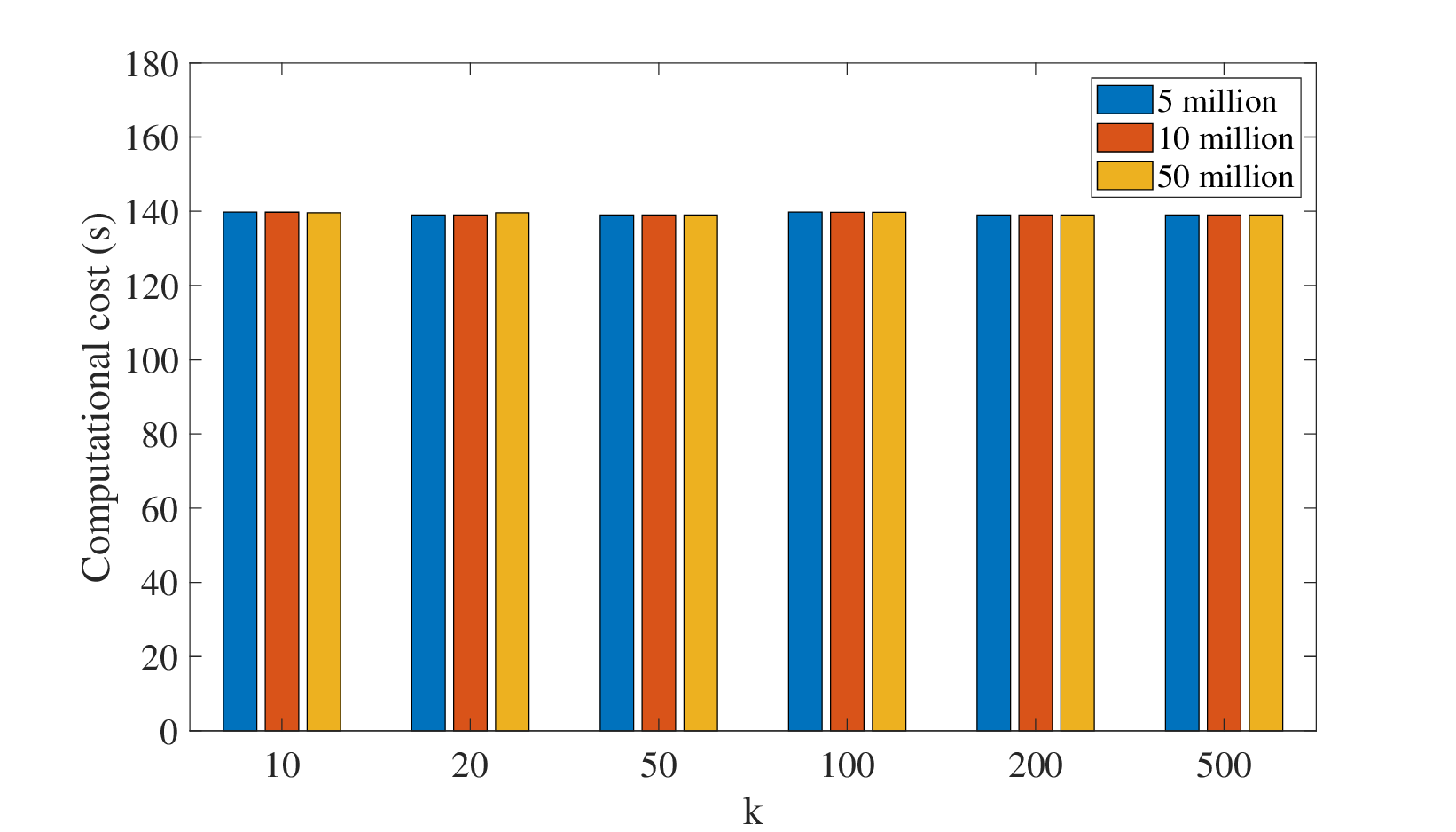}}
\end{minipage}%
\caption{The memory uses of the proposed TLMB and SSMB methods with different $k$ values. (a) TLMB and (b) SSMB.}
\label{fig:mu} 
\end{figure*}

Fig. \ref{fig:cc} shows the computational costs of TLMB and SSMB with different $k$ values. As expected, the computational cost gradually increases as the number of IP records increases from 5 million to 50 million. Moreover, the computational costs of TLMB and SSMB remain relatively stable when $k$ varies from 10 to 500 on each synthetic dataset. This finding demonstrates the stability of TLMB and SSMB for computational efficiency when collecting the statistics of large-scale IP address data. Fig. \ref{fig:mu} shows the memory uses  of TLMB and SSMB with different $k$ values. We observe that TLMB requires more memory use as the number of IP records grows. In contrast, SSMB maintains relatively stable memory across varying numbers of IP records. This finding indicates that SSMB can satisfy certain memory requirements when handling varying numbers of IP records.

\subsection{Discussion}
The gap in computational cost between TLMB and Hash Mapping dramatically increases when the number of IP records increases from 5 million to 50 million. This is because TLMB avoids hash collisions when an IP address is mapped to the corresponding memory block. The computational cost of TLMB is linearly proportional to the number of IP addresses. Moreover, the memory use of SSMB remains almost unchanged regardless of the number of IP addresses. This is consistent with the theory underlying the second proposed mapping mechanism. There is a negligible effect on the computational costs of TLMB and SSMB when $k$ increases from 10 to 100. Moreover, the changes in the memory use of TLMB and computational cost of SSMB are tolerable in practical applications as the number of IP addresses increases. Consequently, TLMB and SSMB reach a reasonable balance between computational cost and memory use when compared with Hash Mapping and IP Mapping. This reveals that the two relationship mapping mechanisms for memory blocks and IP addresses are effective approaches for the design of the four-dimensional sparse matrix.

The memory blocks designed in TLMB and SSMB exhibit superior relationship mapping capabilities compared to those of hash mapping. The memory block takes fully advantages of the inherent property of the memory address, which is employed to corresponding to each part of an IP address. This indicates that integrating the memory block into TLMB and SSMB is both time-stable and memory-stable. In contrast, hash mapping uses a pair of key and value to store the statistics of IP address data. Unfortunately, IP addresses are often sparse in practical applications. Hash mapping requires additional memory to store the remaining two or three parts of the IP address as keys corresponding to TLMB and SSMB, respectively. This has a significant negative impact on the memory use of hash mapping. Therefore, the proposed memory block significantly enhances the capacity of TLMB and SSMB in collecting the statistics of large-scale IP address data.

\section{Conclusion}
\label{sec:Conclusions}
The collection of the statistics of large-scale IP address data is one of the most fundamental problems in network traffic measurement. In this paper, we addressed this problem. Specifically, the two proposed methods present two different relationship mapping mechanisms between memory blocks and IP addresses to strike a balance between computational cost and memory use. They can be employed to search for frequently occurring IP addresses in practical applications. The extensive experimental results demonstrate the effectiveness of the proposed methods.

\bibliographystyle{splncs03}
\bibliography{ess}

\end{document}